\documentclass[aps,pre,twocolumn,showpacs,superscriptaddress,10pt]{revtex4-1}
\usepackage{graphicx}
\usepackage{amsmath}
\usepackage{amsfonts}
\usepackage{amssymb}
\usepackage{amsthm}
\usepackage[colorlinks,citecolor=blue,linkcolor=blue]{hyperref}
\usepackage{color}

\begin{document}

\title{Drift of Charge Carriers in Crystalline Organic Semiconductors}

\author{Jingjuan Dong} \affiliation{State Key Laboratory of Surface Physics and Department of Physics, Fudan University, Shanghai 200433, China}
\author{Wei Si} \affiliation{State Key Laboratory of Surface Physics and Department of Physics, Fudan University, Shanghai 200433, China}
\author{Changqin Wu}\email [Email: ] {cqw@fudan.edu.cn} \affiliation{State Key Laboratory of Surface Physics and Department of Physics, Fudan University, Shanghai 200433, China} \affiliation{Collaborative Innovation Center of Advanced Microstructures, Fudan University, Shanghai 200433, China}

\date{\today}

\begin{abstract}

We investigate the direct-current response of crystalline organic semiconductors in the presence of finite external electric fields by the quantum-classical Ehrenfest dynamics complemented with instantaneous decoherence corrections (IDC). The IDC is carried out in the real-space representation with the energy-dependent reweighing factors to account for both intermolecular decoherence and energy relaxation by which conduction occurs. In this way, both the diffusion and drift motion of charge carriers are described in a unified framework. Based on an off-diagonal electron-phonon coupling model for pentacene, we find that the drift velocity initially increases with the electric field and then decreases at higher fields due to the Wannier-Stark localization, and a negative electric-field dependence of mobility is observed. The Einstein relation, which is a manifestation of the fluctuation-dissipation theorem, is found to be restored in electric fields up to $\sim$10$^5$ V/cm for a wide temperature region studied. Furthermore, we show that the incorporated decoherence and energy relaxation could explain the large discrepancy between the mobilities calculated by the Ehrenfest dynamics and the full quantum methods, which proves the effectiveness of our approach to take back these missing processes.

\end{abstract}


\maketitle

\section{Introduction}

In recent years, the continuous improvement of material quality in functional organic field-effect transistors has largely reduced the extrinsic disorder effects. The ensuing intrinsic behavior has triggered renewed interests in the transport of crystalline organic semiconductors.\cite{Gershenson, Hasegawa} Comprehensive experimental studies have revealed unconventional transport phenomena. The mobility is found to be decreasing with increasing temperature, which is a bandlike behavior.\cite{Podzorov, Sakanoue} Meanwhile, the participating electronic states are found to be localized, which is a sign of incoherent hopping transport. For example, the predicted mean-free-path is comparable or even lower than the intermolecular spacing.\cite{Cheng, Marumoto} It has long been recognized that the electron-phonon interaction, especially the off-diagonal type, plays an important role in such systems. Specifically, the localization is caused by phonon thermal fluctuations, rather than the usual self-trapping mechanism, which supports polarons in conjugated polymers.\cite{Heeger} Furthermore, the dynamic nature of phonon system enable the diffusion of the localized electronic states, leading to finite conduction.\cite{Troisi} This physical picture, termed as transient localization,\cite{Fratini2015} can be described by the mixed quantum-classical Ehrenfest dynamics, in which classical approximation has been imposed on the phonon system. The prediction of this method, such as the temperature dependence of mobility and localization length, are claimed quantitatively in agreement with experimental results.\cite{Troisi2007}

Despite of these successes, there is still a limitation pertaining to the Ehrenfest dynamics. The common procedure for calculating the mobility starts with the evolution of the diffusion dynamics for the diffusion constant, which is then transformed to mobility by evoking the Einstein relation.\cite{Troisi} However, the drift motion under finite electric field, which could be used to calculat the mobility directly, can not be accounted for in the same framework, which will be proven later in this paper. The drift can not be recovered even by considering additional scattering by the stochastic Liouville equation.\cite{Dunlap1986, Dunlap1988} This phenomenon of diffusion dynamics without proper drift motion is in violation of the fluctuation-dissipation theorem and closely related to the drawbacks of the Ehrenfest dynamics. Firstly, the evolution of the electronic state is formally coherent without a specific localization length, which is a sign of the over-coherence that has long been realized.\cite{Landry} Secondly, the energy relaxation is not properly described, resulting in incorrect long-time dynamics. It has been shown that, in the long-time limit, all adiabatic states tends to be equally populated,\cite{Blum} even with the Hellmann-Feynman force.\cite{Parandekar} Also, the validity of the Einstein relation, which is a result of the fluctuation-dissipation theorem and detailed balance, is questionable, since the over-coherence has drawn the system away from the thermal equilibrium. Many efforts have been made to overcome these problems. Ciuchi et al. introduced a phenomenological relaxation time approximation to correct the long-time dynamics.\cite{Ciuchi} Wang and Beljonne generalized the surface hopping approach to such systems.\cite{Tully,WangJPCL} Very recently, two of the present authors proposed an instantaneous decoherence correction (IDC) approach with energy-dependent reweighing factors to account for the decoherence and energy relaxation processes.\cite{Si,Yao} It is shown that this method is able to maintain the near-equilibrium distribution of electronic states in the diffusion dynamics, which is a sign of proper treatment of energy relaxation.

In this paper, we generalize the IDC approach to investigate the drift of charge carriers in finite external electric fields based on the Ehrenfest dynamics within a model of an off-diagonal electron-phonon coupling, which is widely used for the charge transport in crystalline organic semiconductors. This is essential in forming a comprehensive physical picture of carrier dynamics with both diffusion (fluctuation) and drift (dissipation). This approach could also be used to study the electric-field dependence of drift motion, which is an important topic alongside the temperature dependence.\cite{Lee} In many experiments, relatively high fields have been achieved with observation of non-trivial field-dependence of mobility.\cite{Sun} Theoretically, in a perfect 1D system with electric field, the energy eigenstates become localized Wannier-Stark states.\cite{Hartmann,Vidmar,Golez,Cheung} At low electric fields, the semiclassical Blotzmann transport equation\cite{Jacoboni,Wacker,Khan} is well-suited in which the carriers are viewed as classical particles, and the distribution function and current density can be obtained. At higher fields, the Wannier-Stark states become the favorable basis, and the charge carriers drift along by incoherent hopping among different Wannier-Stark states with emission of phonons.\cite{Emin,Rott,Rosini} In the present problem, these physical pictures are not directly applicable due to the additional dynamic disorder, which brings about new localization mechanism (dominant under low fields) and energy distribution. Thus further studies are necessary. This paper is organized as follows. In Sec. II, the quantum-classical dynamics and the IDC method is described. The results of the carrier drift calculated by the present method are shown in Sec. III, together with discussions on the Einstein relation. We conclude this paper with a brief summary in the last section.

\section{Model and method}

We first present the Ehrenfest dynamics of the off-diagonal electron-phonon coupling model for the charge transport in crystalline organic semiconductors. Then the IDC is introduced to incorporate decoherence and energy relaxation in the dynamics.

\subsection{Quantum-classical dynamics}

We consider a one-dimensional molecular chain, which is consisted of $N$ identical molecules abstracted as transport sites. Each site is associated with a vibrational degree of freedom with displacement $u_n$ from its equilibrium position due to the thermodynamic fluctuations, and the transfer integral is modulated by it. For convenience, we choose electron on the LUMO as the charge carrier in the following. The Hamiltonian is similar to the semiclassical Su-Schrieffer-Heeger (SSH) model\cite{Heeger} and is consisted of two parts $H=H_{\text{el}}+H_{\text{la}}$. The electronic part $H_{\text{el}}$ is
\begin{equation}
H_{\text{el}}=\sum_n t_n\left(c_{n+1}^\dagger c_n+c_{n+1}c_n^\dagger\right)-e E \sum_n\left(na+u_n\right)c_n^\dagger c_n,
\end{equation}
where $t_n=-t_0+\alpha\left(u_{n+1}-u_n\right)$; $t_0$ is the transfer integral at equilibrium geometry; $\alpha$ is the electron-phonon coupling strength; $c_n^\dagger\left(c_n\right)$ is the creation (annihilation) operator of electron on molecule $n$; $a$ is the lattice constant; $e$ is the absolute value of the elementary charge; $E$ is the external electric field. The phonon part $H_{\text{la}}$ is
\begin{equation}
H_{\text{la}}=\sum_n\left(\frac{p_n^2}{2m}+\frac{1}{2}m\omega_0^2u_n^2\right),\label{hamiltonian}
\end{equation}
where $\omega_0$ is the frequency of the phonon; $m$ is the molecular mass. The parameters are chosen following that of Troisi and Orlandi for pentacene,\cite{Troisi} which are $t_0=300$ cm$^{-1}$, $\alpha=995$ cm$^{-1}$$\cdot$\AA$^{-1}$, $m=250$ amu, $\omega_0=7.62\times10^{-3}$ fs$^{-1}$, and $a=4$ {\AA}, unless otherwise specified.

In the quantum-classical Ehrenfest dynamics, the electronic system is treated quantum-mechanically and described by the wave function $\left|\psi\left(t\right)\right\rangle$, the evolution of which follows the time-dependent Schr\"{o}dinger equation
\begin{equation}
i\hbar\frac{\partial\left|\psi\left(t\right)\right\rangle}{\partial{t}}=H_{\text{el}}\left|\psi\left(t\right)\right\rangle.\label{wf}
\end{equation}

Because of the large mass of molecule, $m$, the frequency $\omega_0$ is small with $k_BT>\hbar\omega_0$, where $T$ is the temperature, and $k_B$ is the Boltzmann constant. Thus the phonon system can be approximated as classical, and in this way, the evolution of the phonons are governed by the Newton's equation, which is
\begin{equation}
m\ddot{u}_n=-m\omega_0^2u_n-\frac{\partial\left\langle\psi\left(t\right)|H_{\text{el}}|\psi\left(t\right)\right\rangle}{\partial u_n}.\label{lf}
\end{equation}
The initial condition of displacements and velocities satisfy the Maxwell distribution with variance $k_BT/m\omega_0^2$ and $k_BT/m$ respectively. The initial electron state is chosen to be on a single site. The evolution is carried out by the 4th-order Runge-Kutta algorithm, with the integration time step at most 0.25 $\hbar/t_0$, which is sufficiently small to not influence the final results. A chain with $N=400$ molecules is used with fixed boundary condition and the results are got by averaging over 50000 realizations.

\subsection{Instantaneous Decoherence Correction}

As we have discussed before, the Ehrenfest dynamics alone cannot properly account for the decoherence and energy relaxation processes, which are key to the proper drift that is studied in this work. For this purpose, we employ the IDC technique to account for these processes. The main idea of IDC is to incorporate decoherence by random projection of electronic wave function in a certain basis controlled by the decoherence time $t_d$. The energy relaxation can be further considered by energy-dependent reweighing factors in the projection. Similar approaches have been used in the Anderson model and spin dynamics of excited states.\cite{Flores,Kominis} It is recently applied to the crystalline organic semiconductors, resulting in a diffusion with localized states and band-like behavior.\cite{Yao} Besides, the near-equilibrium Boltzmann distribution is maintained for the electronic system.\cite{Si} Here we generalize this method to account for the drift motion in finite electric fields, which is implemented in site representation with Miller-Abrahams type energy-dependent reweighing factor.\cite{Miller} The Miller-Abrahams rate is widely used to study the charge transport in amorphous organic semiconductors, and satisfies the detailed balance requirement, which is essential for proper energy relaxation.\cite{Bassler}

Specifically, the IDC is implemented by Monte Carlo algorithm and the detailed procedure is as follows. Initially, the electron is placed on a single molecule. Then the system is evolved using Ehrenfest dynamics, following Eqs.~(\ref{wf}) and (\ref{lf}) for a time interval $t_d$. The electron wave function, $\left|\psi\left(t\right)\right\rangle$ is expressed in Wannier representation as $\left|\psi\left(t\right)\right\rangle=\sum_{n}\varphi_n\left(t\right)\left|n\right\rangle$.
After the evolution, the electron spreads onto several molecules, and the population on molecule $n$ is $|\varphi_n|^2$. At this point, a projection operation is imposed for the electronic wave function. The electron is either projected onto the initial molecule, or transits to another molecule. The transition rate from original molecule $m$ to target molecule $n$ is determined by the product of the electron population on every molecule and the Miller-Abrahams type energy-dependent reweighing factor, which is
\begin{equation}
g_{m\rightarrow n}=\left\{
\begin{array}{cc}
C_n|\varphi_n|^2\exp\left[-\left(\varepsilon_n-\varepsilon_m\right)/k_BT\right], & \varepsilon_n\geq\varepsilon_m,\\
C_n|\varphi_n|^2, & \varepsilon_n<\varepsilon_m,
\end{array}
\right.\label{rate}
\end{equation}
where $C_n$ is a normalization constant, and $\varepsilon_n=\left(na+u_n\right)E$ is the electric potential energy on molecule $n$. A random number, $\chi$, uniformly distributed between 0 and 1, is generated to select the target molecule of the projection. Then a new round of evolution is started.

Here, we would like to discuss the choice of decoherence time, which is an important parameter in our study. The experimental determination of this parameter is not easy. Many theoretical studies have been contributed to this topic.\cite{Jaeger} The decoherence process is often simulated by the decay rate of the off-diagonal elements of the reduced density matrix.\cite{Hwang,Madrid} In some cases, the decoherence time can be roughly estimated by the intermolecular overlap integral, i.e., $t_d\sim \hbar/t_0$, which can be confirmed by the calculation of purity in spin-boson model.\cite{Paganelli} Many semiclassical theories have also been proposed for the decoherence time in electron-phonon coupled systems,\cite{Neria} which can be evaluated by the difference between forces on diverging classical trajectories\cite{Bedard} or the potential energy difference associated with different adiabatic states.\cite{Zhu} Based on different methods and approximation, the resulting decoherence time ranges from a few to hundreds of femtoseconds. Using a short time approximation, a simple formula of decoherence time is derived based on the decay rate of the energy correlation function, which is $t_d=\hbar/\sqrt{\lambda k_BT}$.\cite{WangJCP} According to this function, the decoherence time is about $\sim10$ fs (the reorganization energy $\lambda=159$ meV\cite{Troisi2007} is used) for rubrene, and $\sim$12 fs (the reorganization energy $\lambda=118$ meV\cite{Gruhn} is used) for pentacene at room temperature. In our method, the decoherence time is taken to be a specific value independent of temperature. We investigat the decoherence time ranging from 10 fs to 100 fs, and $t_d=10$ fs is used generally, unless otherwise specified.

\section{RESULTS AND DISCUSSION}

In this section, we present the results obtained by applying the IDC approach to the off-diagonal electron-phonon coupling model. We first calculate the drift velocity of charge carriers and the drift mobility, and then move on to discuss the Einstein relation.

\subsection{Drift of charge carriers}

We first study the electron drift dynamics with the method introduced above. The dynamical evolution of $N_s$ realizations for the averaged population of electron on different sites $n$ at time $t$ are calculated as
\begin{equation}
P_n\left(t\right)=\frac{1}{N_s}\sum\limits_{s=1}^{N_s}\left|\varphi_n^s\left(t\right)\right|^2,
\end{equation}
where $s$ is the index for realization. The displacement of the electron is
\begin{equation}
\Delta x\left(t\right)=a\sum\limits_{n=1}^{N}nP_n\left(t\right)-x\left(0\right),
\end{equation}
where $x\left(0\right)$ is the initial position of the electron. The drift velocity $v_d$ is calculated at long-enough times when $\Delta x(t)$ increases linearly, which is
\begin{equation}
v_d=\lim_{t\rightarrow+\infty}\frac{\partial\Delta x\left(t\right)}{\partial t}\label{velocity}.
\end{equation}

\begin{figure}
\includegraphics{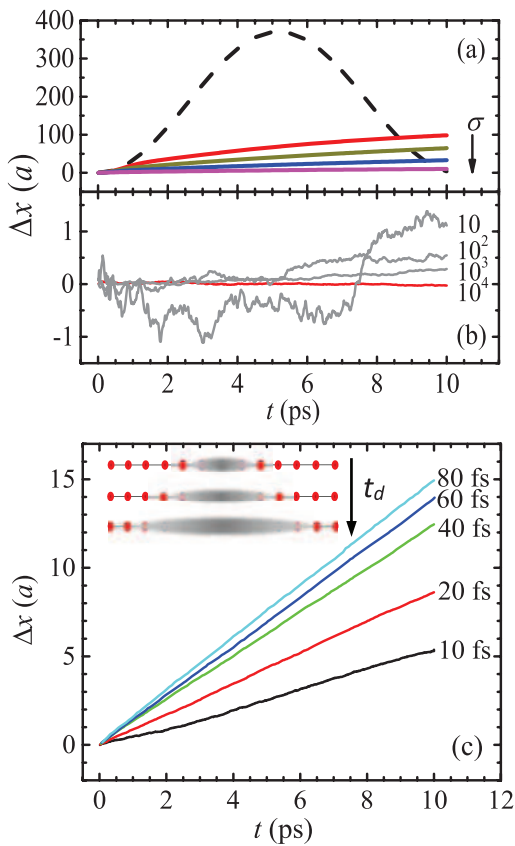}
\caption{(Color online) (a) Displacement versus time in the Ehrenfest dynamic with a Gaussian wave packet. BOs in perfect periodic lattice is shown (dash black line). The solid lines are displacement with lattice displacement disorder, $\sigma$, which increases as shown by the arrow. (b) Displacement versus time of the single-site occupied electron (red line). The results with 10, 100 and 1000 realizations are also shown (grey solid line). (c) Displacement versus time with different decoherence times at the temperature $T=150$ K. Inset: Scheme of the distribution of the electron with increasing decoherence time. For all panels, the electric field $E=10^4$ V/cm.}
\label{fig-01}
\end{figure}

We first prove the absence of the electron drift motion with the Ehrenfest dynamics in the dynamic disorder regime. For this purpose, we calculat the dynamics with two representative initial conditions of the electron wave function. The first one features a Gaussian wave packet with a central momentum $p_0=0$; in the second one the electron is localized on a single site. We present the results of $\Delta x(t)$ with  different magnitude of disorder tuned by the temperature, which is shown in Fig.~\ref{fig-01}(a). 
When $T=0$, the system is a perfect one-dimensional lattice without dynamic disorder. With an applied static electric field, the eigenstates are localized Wannier-Stark states and the energy spectrum is consisted of Wannier-Stark ladders. The current response behaves as the well-known Bloch oscillations (BOs),\cite{Hartmann} which has been observed in a variety of experiments.\cite{Waschke,Dahan} For the Gaussian case, the center of the wave packet oscillates periodically, as is shown by the dashed line in Fig.~\ref{fig-01}(a); and for the single-site case, the BOs is manifested as a breathing mode, i.e., the wave packet widens and shrinks periodically with a fixed center. In both cases, although the momentum is accelerated by the electric field, it changes direction when it reaches the boundary of the Brillouin zone and no direct-current response is present. With the dynamic disorder at finite temperatures, the current response is different. For the Gaussian case, only a slight displacement is observed as shown in Fig.~\ref{fig-01}(a). However, the displacement gradually stops in the long time limit. Thus the motion in this case is still not a proper drift. For the single-site case, the wave packet is not guaranteed to move along the electric field and $\Delta x(t)$ tends to zero with increasing number of realizations, which is shown in Fig.~\ref{fig-01}(b). So there is no drift motion in this case either. Further, it can be anticipated that no direct-current response can be observed whatever the initial state is.\cite{Parandekar,Si} In all, the Ehrenfest dynamics alone can not describe the direct-current response anticipated with finite external electric field.

We then move on to the results with the IDC method, and the calculated $\Delta x(t)$ at different $t_d$ is shown in Fig.~\ref{fig-01}(c). It can be seen that the displacement increases linearly with time at large times for all the decoherence times considered, which means a drift motion driven by the external electric field resulting in a steady electric current. Moreover, the electron drifts faster with increasing decoherence time. This phenomenon is due to the disturbance of the electronic dynamics by the IDC. With decreasing $t_d$, the population is limited on fewer molecules which is similar to the quantum Zeno effect,\cite{Itano} resulting in slower drift as is shown by the inset in Fig.~\ref{fig-01}(c).

\begin{figure}
\includegraphics{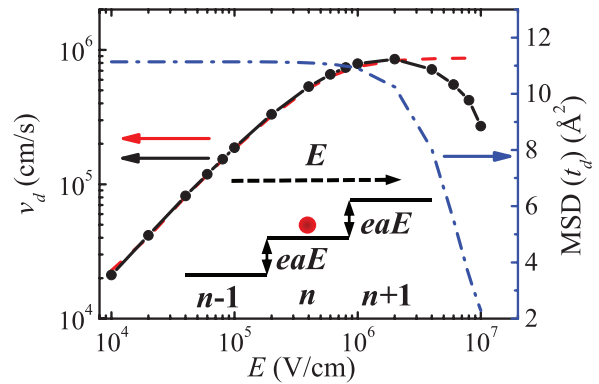}
\caption{(Color online) Electric field dependence of drift velocity (left scale) and MSD (right scale) at $T=150$ K. The black circle line is the drift velocity calculated in our model. The red dashed line is the drift velocity obtained in the model, where the evolution of the wave function is not affected by the electric field, which is only incorporated in decoherence process. The blue dot dashed line shows MSD at $t_d=10$ fs. Inset: Scheme of the electronic energy in the equidistant three-molecule model. The direction of the electric field is on the right. The electron is placed on the middle site, and the energy difference between adjacent sites is $eaE$.}
\label{fig-02}
\end{figure}

We further calculated the electric-field dependence of drift velocity $v_d$ for $t_d=10$ fs and $T=150$ K, which is shown in Fig.~\ref{fig-02}.
It can be seen that $v_d$ first increases with increasing electric field. Then $v_d$ gradually levels off as the difference of electric potential energy between nearest molecules is comparable to $k_BT$ ($E\sim3.2\times10^5$ V/cm with the parameter used), and then reaches its maximum value. With further increase of the electric field, $v_d$ began to fall rapidly beyond $E\sim2\times10^6$ V/cm. Higher electric-field strength would be beyond realistic situations. For pentacene, the fields for substantial decrease of $v_d$ is too high to be realized. Besides in functional devices, the electric field dependence is further complicated by other factors, such as the increase of `mobile' carrier density.\cite{Podzorov} These may be the reasons why this trend is not observed.

For a better understanding of the electric-field dependence of $v_d$, two cases are studied, in which the roles of electric field are investigated separately in two processes. In the first case, the electric field is removed from the Ehrenfest dynamics and is present in the IDC; while in the second one, the electric field is maintained in the Ehrenfest dynamics and is removed from the IDC. For the first case, a proper electron drift is still present with the resulting $v_d$ shown by the dash line in Fig.~\ref{fig-02}. The electric-field dependence of $v_d$ in this case can be analyzed in the limit in which only the nearest molecules are populated by the Ehrenfest dynamics. A simplified equidistant three-molecule model (inset of Fig.~\ref{fig-02}) can be used with the same adjacent energy difference $eaE$. According to Eq.~(\ref{rate}), the energy-dependent reweighing factor from site $n$ to $n-1$ is 1, which is independent of the electric field, while the factor from site $n$ to $n+1$ is $\exp(-eaE/k_BT)$. Without the electric field in the dynamics, the averaged population should be symmetric. So when $eaE$ is smaller than $k_BT$, the probability of electron transporting downward in energy increases with electric field due to the energy factor in IDC, which leads to a larger $v_d$. When $eaE$ is of the same order as $k_BT$, the increase would slow down and $v_d$ tends to a saturation value. Thus the increase and saturation of $v_d$ comes from the stronger energy asymmetry brought by the electric field. Similar phenomenon has been found in the hopping transport\cite{Jansson} and the dynamics of the polarons.\cite{Ono,Arikabe} However, the ability of electron motion can still be reflected by the mean-square displacement for diffusion, which is calculated as
\begin{equation}
\text{MSD}\left(t\right)=a^2\left\{\sum\limits_{n=1}^{N}n^2P_n\left(t\right)-\left[\sum\limits_{n=1}^{N}n P_n\left(t\right)\right]^2\right\}.
\end{equation}
We calculated the MSD at $t_d$ for the influence of the electric field on the electron dynamics, which is shown by the dotted-dash line in Fig.~\ref{fig-02}. It can be seen that the MSD remains a definite value at low fields and begins to fall rapidly with electric fields around $E=5\times10^5$ V/cm. The increasing potential energy difference between different sites induces localization, which suppresses the diffusion. This decrease is a manifestation of quantum coherent effect and have the same origin as that observed in superlattice.\cite{Kirchoefer,Yang}

\subsection{Drift mobility}

\begin{figure}
\includegraphics{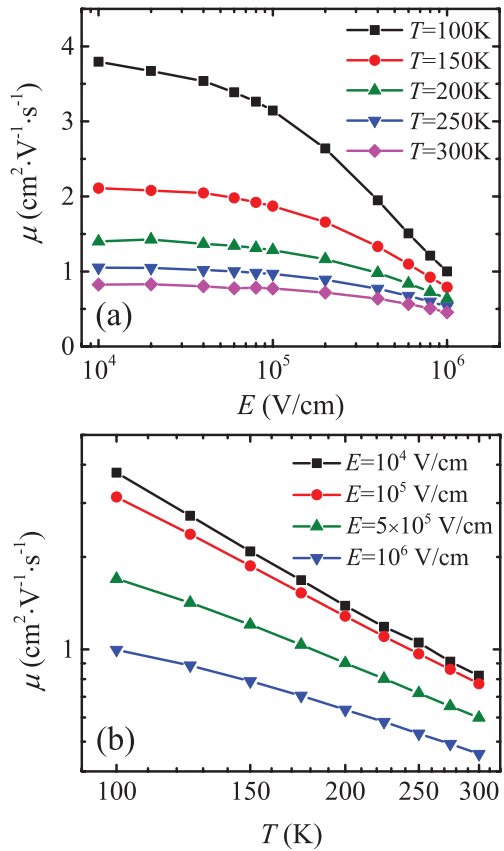}
\caption{(Color online) (a) Electric field dependence of mobility at different temperatures. (b) Temperature dependence of mobility at different electric fields.}
\label{fig-03}
\end{figure}

With the proper drift motion, the carrier mobility can be calculated directly with $\mu=v_d/E$, without invoking the Einstein relation, which is an advantage of our approach. The dependence of mobility on both electric field and temperature are shown in Fig.~\ref{fig-03}.
Firstly, it can be seen that the mobilities obtained are around $\thicksim$1 cm$^2\cdot$V$^{-1}\cdot$s$^{-1}$, which are in agreement with the available experimental measurements for pentacene.\cite{Lin} The electric-field dependence of mobility is shown in Fig.~\ref{fig-03}(a). Under lower field, the mobility tends to a saturate value with decreasing field. The increase of drift velocity $v_d$ with field become sub-linear at finite field, leading to the gradual decrease of mobility. Under higher fields, the $v_d$ itself begin to decrease due to the localization mechanism presented above. Thus the mobility decreases more steeply.

Experimentally, the negative dependence is observed in ultrapurified naphthalene single crystals at lower temperature ($<100$ K).\cite{Warta} In functional field-effect transistors, hopping-like behavior is commonly observed with increasing mobility with field, which can be fitted with either the Fowler-Nordheim or Poole-Frenkel line shape. The negative dependence is ascribed to the fact that the present model is more suitable for the intrinsic transport in single-crystals. The mobility from field-effect devices is further affected by extrinsic defects, particularly at the gate interface. These could act as shallow traps, which hinder the quantum transport and influence the electric-field dependence. Under low temperature and in single crystals, the molecular arrangements are highly ordered, which makes the quantum effect predominant.\cite{Warta} It is worth to motion that, very few organic molecules can be grown into single crystals without grain boundaries, and the crystal growth process is long and costly. So the negative dependence is hard to be observed in field-effect devices.\cite{kaake} It would be meaningful to discuss the situation with shallow traps and the relevant study is under progress.

We also calculate the temperature dependence of mobility, which is shown in Fig.~\ref{fig-03}(b). At the various electric fields considered, it all shows a band-like behavior with power-law dependence $\mu\propto T^{-m}$.\cite{Jurchescu} However, the value of the $m$ index is different, which decreases slightly with increasing field. For example, the best fits give that  $m=1.39$ for $E=10^4$ V/cm, 1.28 for $E=10^5$ V/cm, 0.96 for $E=5\times10^5$ V/cm and 0.73 for $E=10^6$ V/cm. In all, the electric-field dependence is less sensitive with increasing fields, as the field itself is becoming more important in the carrier dynamics compared to the influence of the thermal fluctuation.

\subsection{Einstein relation}

\begin{figure}
\includegraphics{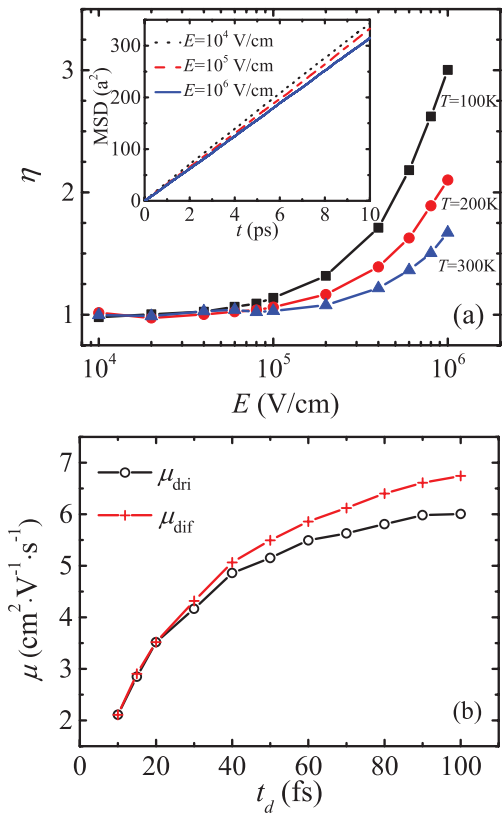}
\caption{(Color online) (a) Ratio between diffusion coefficient and mobility versus the electric field at different temperatures. Inset: Time dependence of MSD with different electric fields at $T=150$ K. The decoherence time $t_d=10$ fs. (b) Drift mobility (black circles) and mobility calculated with Einstein relation (red crosses) versus decoherence time. $T=150$ K, $E=10^4$ V/cm.}
\label{fig-04}
\end{figure}

We further explore the Einstein relation, which is the manifestation of the fluctuation-dissipation theorem, by calculating the ratio between the diffusion coefficient and the mobility, $\eta=eD/\mu k_BT$. The diffusion coefficient is calculated with
\begin{equation}
D=\frac{1}{2}\lim_{t\rightarrow+\infty}\frac{\partial \text{MSD}\left(t\right)}{\partial t}.
\end{equation}
The time evolution of MSD at $T=150$ K is shown in the inset of Fig.~\ref{fig-04}(a) with three representative electric fields. 
All results begin to increase linearly with time after a short initial period, leading to a well-defined diffusion coefficient. Combined with the mobility from drift, the electric-field dependence of the ratio $\eta$ is shown in Fig.~\ref{fig-04}(a), with $T=100$ K, 200 K and 300 K as representative temperatures. At relatively low fields, the ratio tends to 1, as is required by the Einstein relation. The restoring of the Einstein relation for fields $<10^5$ V/cm is a signature of the effectiveness of our approach, which means the proper energy relaxation obeying the fluctuation-dissipation theorem. With larger electric fields, the ratio increases rapidly to nearly 3 at $E=10^6$ V/cm for $T=100$ K, which is due to the deviation from the near-equilibrium situation. It also becomes temperature-dependent, with larger $\eta$ for lower temperature.

We further studied the dependence on decoherence time, which is shown in Fig.~\ref{fig-04}(b). Two kinds of mobility are shown for comparison: $\mu_{\rm dri}$ from the drift and $\mu_{\rm dif}$ from the diffusion coefficient with the Einstein relation. Both kinds of mobility increases with $t_d$ yet the ratio remains close to 1. In detail, $\mu_{\rm dif}$ increases faster, resulting in a slight increase of $\eta$ from 1.0 at $t_d=10$ fs to 1.12 at $t_d=100$ fs. The difference can be ascribed to the less effective account of energy relaxation by IDC with increasing $t_d$. However we note that even with a decoherence time $t_d$ as large as 100 fs, the deviation is still small ($\eta=1.12$). In the extreme case of very large $t_d$, the evolution returns to the usual Ehrenfest dynamics. Then there would be no drift motion and the ratio would tend to infinity.

\subsection{Comparison with other approaches}

\begin{figure}
\includegraphics{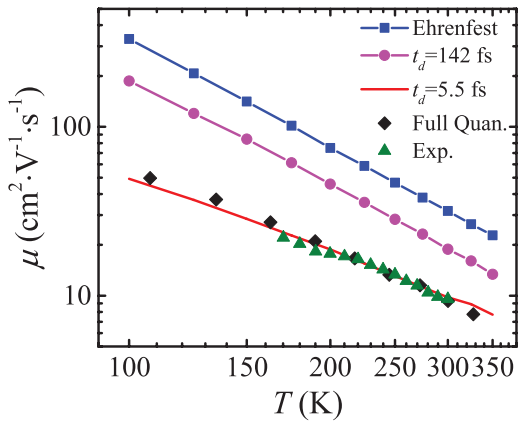}
\caption{(Color online) Temperature dependence of the mobility for rubrene\cite{Podzorov} (green up triangles) compared with our results in small electric field, $E=10^4$ V/cm, with the decoherence time $t_d=142$ fs (purple circle line) and $t_d=5.5$ fs (red line), Ehrenfest dynamics\cite{Troisi} (blue square line) and full quantum dynamics\cite{Filippis} (black diamonds). }
\label{fig-05}
\end{figure}

To further validate our approach, we compare our temperature dependence of mobility with both the experimental values\cite{Podzorov} and that by the full quantum approach.\cite{Filippis} The results are shown in Fig.~\ref{fig-05} with the parameters following De Filippis et al. for rubrene,\cite{Filippis} which are $t_0=0.093$ eV, $\alpha=0.29$ eV$\cdot${\AA}$^{-1}$, $m=532$ amu, $\omega_0=7.06\times10^{-3}$ fs$^{-1}$ and $a=7.2$ {\AA}.
It can be seen that a fairly good comparison is achieved for the temperature region considered when $t_d$ is taken as 5.5 fs. Here the experimental values are extracted from the Hall effect in single-crystal field-effect transistors by Podzorov et al.; the mobilities of full quantum approach are calculated by the linear response theory and quantum Monte Carlo method. We also present the mobility values calculated by the Ehrenfest dynamics, which are almost one order of magnitude larger than the experimental values. This large discrepancy can be ascribed to the decoherence and energy relaxation processes that are not properly treated in the Ehrenfest dynamics, which can be effectively taken back by the IDC approach proposed here.

We would like to stress that the above comparison is achieved by tuning only one parameter, which is the decoherence time. The value of 5.5 fs is on the same order of magnitude with the estimated one of 10 fs mentioned previously.\cite{WangJCP} We also show the results with the decoherence time equaling the characteristic time of phonon system, i.e., $t_d=1/\omega_0=142$ fs, where $\omega_0$ is the phonon frequency. The resulting mobility also shows deviation from the experimental value, which follows the trend of increasing mobility with $t_d$ shown in Fig.~\ref{fig-01}(b). This contrast shows that the smaller $t_d$, comparable to the characteristic time scale of transfer integral, is more reasonable, under which the Einstein relation is also better obeyed.

In the end, we would like to add some comments on the comparison of the present approach with the other ones based on quantum-classical approximation. A very promising one is the relaxation time approximation proposed by Mayou et al.,\cite{Ciuchi} in which the velocity correlation function is corrected for calculating mobility using the linear response theory. Besides, the present approach shares similar opinion with Zuppiroli et al.\cite{Picon} in that decoherence is essential for conduction. In their treatment, the acoustic phonon velocity is used as the average velocity of carrier. Both approaches starts with the diffusion and the mobility is calculated by the Einstein relation. The drift is not explicitly studied, which is the topic of the present work. The resulting carrier dynamics is a long-time successive motion in real space, and the mobility is directly calculated from the drift velocity. We have also obtained a negative temperature dependence of mobility like that of the two approaches.

\section{CONCLUSION}

In summary, we investigate the drift motion of crystalline organic semiconductors in a finite electric field by the mixed quantum-classical Ehrenfest dynamics with decoherence correction. The decoherence is implemented by projection operations and the energy relaxation is considered by energy-dependent reweighing factor. These processes are key to a comprehensive picture of carrier dynamics including both diffusion and drift. Applying to an off-diagonal electron-phonon coupling model, our approach successfully describes the drift motion of charge carrier. We find that the drift velocity increases with external field at low field range; at high electric fields the drift velocity begin to decrease due to Wannier-Stark localization. The value of the resulting mobilities are about $\sim1$ cm$^2\cdot$V$^{-1}\cdot$s$^{-1}$, which are in agreement with experimental results. The mobility tends to a saturate value at low fields and decreases at large fields. Furthermore, the present approach could well restore the Einstein relation at low electric fields up to $\sim10^5$ V/cm with decoherence times of 10$\sim$20 fs. Finally, the mobility from our approach compares well with both the experimental value and that by the full quantum techniques. The implemented $t_d$ of 5.5 fs is on the same order of magnitude as theoretical estimations. Further, the large discrepancy of mobility value from Ehrenfest dynamics and the experimental measurements can be ascribed to the decoherence and energy relaxation that is incorporated by our approach.

\begin{acknowledgments}
We acknowledge the financial supports from the National Natural Science Foundation of China (No. 11574050) and the National Basic Research Program of China (No. 2012CB921402).
\end{acknowledgments}


\begin{thebibliography}{99}
\bibitem{Gershenson} M. E. Gershenson, V. Podzorov, and A. F. Morpurgo, Rev. Mod. Phys. {\bf78}, 973 (2006).
\bibitem{Hasegawa} T. Hasegawa and J. Takeya, Sci. Technol. Adv. Mat. {\bf10}, 024314 (2009).
\bibitem{Podzorov} V. Podzorov, E. Menard, J. A. Rogers, and M. E. Gershenson, Phys. Rev. Lett. {\bf95}, 226601 (2005).
\bibitem{Sakanoue} T. Sakanoue and H. Sirringhaus, Nat. Mater. {\bf9}, 736 (2010).
\bibitem{Cheng} Y. C. Cheng, R. J. Silbey, D. A. da Silva Filho, J. P. Calbert, J. Cornil, and J. L. Br\'{e}das, J. Chem. Phys. {\bf118}, 3764 (2003).
\bibitem{Marumoto}K. Marumoto, S. Kuroda, T. Takenobu, and Y. Iwasa, Phys. Rev. Lett. {\bf97}, 256603 (2006).
\bibitem{Heeger} A. J. Heeger, S. Kivelson, J. R. Schrieffer, and W. P. Su, Rev. Mod. Phys. {\bf60}, 781 (1988).
\bibitem{Troisi} A. Troisi and G. Orlandi, Phys. Rev. Lett. {\bf96}, 086601 (2006).
\bibitem{Fratini2015} S. Fratini, D. Mayou, and S. Ciuchi, arXiv:1505.02686v1 (2015).
\bibitem{Troisi2007} A. Troisi, Adv. Mater. {\bf19}, 2000 (2007).
\bibitem{Dunlap1986} D. H. Dunlap and V. M. Kenkre, Phys. Rev. B {\bf34}, 3625 (1986).
\bibitem{Dunlap1988} D. H. Dunlap and V. M. Kenkre, Phys. Rev. B {\bf37}, 6622 (1988).
\bibitem{Landry} B. R. Landry and J. E. Subotnik, J. Chem. Phys. {\bf137}, 22A513 (2012).
\bibitem{Blum} K. Blum, Density Matrix Theory and Applications, 275 (Plenum, New York, 1996).
\bibitem{Parandekar} P. V. Parandekar and J. C. Tully, J. Chem. Phys. {\bf122}, 094102 (2005).
\bibitem{Ciuchi} S. Ciuchi, S. Fratini, and D. Mayou, Phys. Rev. B {\bf83}, 081202 (2011).
\bibitem{Tully} J. C. Tully, J. Chem. Phys. {\bf93}, 1061 (1990).
\bibitem{WangJPCL} L. Wang and D. Beljonne, J. Phys. Chem. Lett. {\bf4}, 1888 (2013).
\bibitem{Si} W. Si and C. Q. Wu, J. Chem. Phys. {\bf143}, 024103 (2015).
\bibitem{Yao} Y. Yao, W. Si, X. Hou, and C. Q. Wu, J. Chem. Phys. {\bf136}, 234106 (2012).
\bibitem{Lee} J. Lee, J. W. Chung, D. H. Kim, B. L. Lee, J. I. Park, S. Lee, R. Hausermann, B. Batlogg, S. S. Lee, I. Choi, I. W. Kim, and M. S. Kang, J. Am. Chem. Soc. {\bf137}, 7990 (2015).
\bibitem{Sun} H. Sun, Q. Wang, Y. Li, Y. F. Lin, Y. Wang, Y. Yin, Y. Xu, C. Liu, K. Tsukagoshi, L. Pan, X. Wang, Z. Hu, and Y. Shi, Sci. Rep. {\bf4}, 7227 (2014).
\bibitem{Hartmann} T. Hartmann, F. Keck, H. J. Korsch, and S. Mossmann, New J. Phys. {\bf6}, 2 (2004).
\bibitem{Vidmar} L. Vidmar, J. Bon\v{c}a, M. Mierzejewski, P. Prelov\v{s}ek, and S. A. Trugman, Phys. Rev. B {\bf83}, 134301 (2011).
\bibitem{Golez} D. Gole\v{z}z, J. Bon\v{c}a, L. Vidmar, and S. A. Trugman, Phys. Rev. Lett. {\bf109}, 236402 (2012).
\bibitem{Cheung} A. K. C. Cheung and M. Berciu, Phys. Rev. B {\bf88}, 035132 (2013).
\bibitem{Jacoboni} C. Jacoboni and L. Reggiani, Rev. Mod. Phys. {\bf55}, 645 (1983).
\bibitem{Wacker} A. Wacker, A. Jauho, S. Rott, A. Markus, P. Binder, and G. H. D\"{o}hler, Phys. Rev. Lett. {\bf83}, 836 (1999).
\bibitem{Khan} F. S. Khan, J. H. Davies, and J. W. Wilkins, Phys. Rev. B {\bf36}, 2578 (1987).
\bibitem{Emin} D. Emin and C. F. Hart, Phys. Rev. B {\bf36}, 2530 (1987).
\bibitem{Rott} S. Rott, N. Linder, and G. H. D\"{o}hler, Phys. Rev. B {\bf65}, 195301 (2002).
\bibitem{Rosini} M. Rosini and L. Reggiani, Phys. Rev. B {\bf72}, 195304 (2005).
\bibitem{Flores}J. C. Flores, Phys. Rev. B {\bf69}, 012201 (2004).
\bibitem{Kominis} I. K. Kominis, New J. Phys. {\bf15}, 075017 (2013).
\bibitem{Miller} A. Miller and E. Abrahams, Phys. Rev. {\bf120}, 745 (1960).
\bibitem{Bassler} H. B\"{a}ssler, Phys. Stat. Sol. (b) {\bf175}, 15 (1993).
\bibitem{Jaeger} H. M. Jaeger, S. Fischer, and O. V. Prezhdo, J. Chem. Phys. {\bf137}, 22A545 (2012).
\bibitem{Hwang} H. Hwang and P. J. Rossky, J. Chem. Phys. {\bf120}, 11380 (2004).
\bibitem{Madrid} A. B. Madrid, K. Hyeon-Deuk, B. F. Habenicht, and O. V. Prezhdo, ACS Nano {\bf3}, 2487 (2009).
\bibitem{Paganelli} S. Paganelli and S. Ciuchi, J. Phys.: Condens. Matter {\bf20}, 235203 (2008).
\bibitem{Neria} E. Neria and A. Nitzan, J. Chem. Phys. {\bf99}, 1109 (1993).
\bibitem{Bedard} M. J. Bedard-Hearn, R. E. Larsen, and B. J. Schwartz, J. Chem. Phys. {\bf123}, 234106 (2005).
\bibitem{Zhu} C. Zhu, A. W. Jasperand, and D. G. Truhlar, J. Chem. Phys. {\bf120}, 5543 (2004).
\bibitem{WangJCP} L. Wang and D. Beljonne, J. Chem. Phys. {\bf139}, 064316 (2013).
\bibitem{Gruhn} N. E. Gruhn, D. A. da Silva Filho, T. G. Bill, M. Malagoli, V. Coropceanu, A. Kahn, and J.-L. Br\'{e}das, J. Am. Chem. Soc. {\bf124}, 7918 (2002).
\bibitem{Waschke} C. Waschke, H. G. Roskos, R. Schwedler, K. Leo, H. Kurz, and K. K\"{o}hler, Phys. Rev. Lett. {\bf70}, 3319 (1993).
\bibitem{Dahan} M. B. Dahan, E. Peik, J. Reichel, Y. Castin, and C. Salomon, Phys. Rev. Lett. {\bf76}, 4508 (1996).
\bibitem{Itano} W. M. Itano, D. J. Heinzen, J. J. Bollinger, and D. J. Wineland, Phys. Rev. A {\bf41}, 2295 (1990).
\bibitem{Jansson} F. Jansson, A. V. Nenashev, S. D. Baranovskii, F. Gebhard, and R. \"{O}sterbacka, Phys. Stat. Sol. (a) {\bf207}, 613 (2010).
\bibitem{Ono} Y. Ono and A. Terai, J. Phys. Soc. Jpn. {\bf59}, 2893 (1990).
\bibitem{Arikabe} Y. Arikabe, M. Kuwabara, and Y. Ono, J. Phys. Soc. Jpn. {\bf65}, 1317 (1996).
\bibitem{Kirchoefer} S. W. Kirchoefer, R. Magno, and J. Comas, App. Phys. Lett. {\bf44}, 1054 (1984).
\bibitem{Yang} S. Y. Yang, P. Liu, S. W. Guo, L. Zhang, D. Yang, Y. R. Jiang, and B. S. Zou, Appl. Phys. Lett. {\bf104}, 033301 (2014).
\bibitem{Lin} Y. J. Lin, H. Y. Tsao, and D. S. Liu, Appl. Phys. Lett. {\bf101}, 013302 (2012).
\bibitem{Warta} W. Warta and N. Karl, Phys. Rev. B {\bf32}, 1172 (1985).
\bibitem{kaake} L. G. Kaake, P. F. Barbara, and X. Y. Zhu, J. Phys. Chem. Lett. {\bf1}, 628 (2010).
\bibitem{Jurchescu} O. D. Jurchescu, J. Baas, and T. T. M. Palstra, Appl. Phys. Lett. {\bf84}, 3061 (2004).
\bibitem{Filippis}G. De Filippis, V. Cataudella, A. S. Mishchenko, N. Nagaosa, A. Fierro, and A. de Candia, Phys. Rev. Lett. {\bf114}, 086601 (2015).
\bibitem{Picon} J. D. Picon, M. N. Bussac, and L. Zuppiroli, Phys. Rev. B {\bf75}, 235106 (2007).
\end{thebibliography}

\end{document}